\documentclass[11pt]{elsart}
\usepackage[dvips]{graphicx,overpic}

\begin{document}
\begin{frontmatter}

\title{Identifying Influential Spreaders by Weighted LeaderRank}

\author{Qian Li$^{a,b,c}$}
\author{Tao Zhou$^{a}$}
\author{Linyuan L\"{u}$^{d,e}$}
\author{Duanbing Chen$^{a}$}

\address
{$^{a}$ Web Sciences Center,University of Electronic Science and
Technology of China,Chengdu 611731, People's Republic of China}

\address
{$^{b}$ Big Data Lab, Cloud Valley, Beijing 100176, People's Republic of China}

\address
{$^{c}$ College of Physics and Technology, Guangxi Normal
University, Guilin 541004, People's Republic of China}

\address
{$^{d}$ Institute of Information Economy, Alibaba Business College, Hangzhou Normal University, Hangzhou 310036, People's Republic of China}

\address
{$^{e}$ Department of Physics, University of Fribourg, Fribourg CH-1700, Switzerland}

\begin{abstract}
Identifying influential spreaders is crucial for understanding and controlling spreading processes on social networks. Via assigning degree-dependent weights onto links associated with the ground node, we proposed a variant to a recent ranking algorithm named LeaderRank [L. L\"u \emph{et al.}, PLoS ONE 6 (2011) e21202]. According to the simulations on the standard SIR model, the weighted LeaderRank performs better than LeaderRank in three aspects: (i) the ability to find out more influential spreaders, (ii) the higher tolerance to noisy data, and (iii) the higher robustness to intentional attacks.
\end{abstract}

\date{}
\end{frontmatter}


\section{Introduction}

The spreading processes of epidemic and information attract increasing attention in complex network studies \cite{Zhou2006}, and researchers tried to find the reason why information spread so quickly \cite{Lu2011,Benjamin2012} as well as how to decelerate the spreading \cite{Markus2012}. Among many ingredients for quick and wide spreading, influential spreaders play a major role \cite{Sinan2012,Renato2012,Borge2012}. Accordingly, immunization on large-degree nodes (they are usually considered to be more influential) is a highly efficient method to control epidemic spreading \cite{Pastor2002,Bai2007,Laurent2012}. It is of great theoretical and practical significance to identify influential spreaders in networks, and similar methods can be applied in ranking scientists \cite{Filippo2009,YanBo2012}, publications \cite{YanBo2012}, athletes \cite{Juyong2005} and finding influential directors \cite{Xu2011}.

How to identify influential spreaders effectively and efficiently is a big challenge up to now. A number of centrality indices have been proposed to address this problem, such as degree centrality, closeness centrality \cite{Sabidussi1966}, betweenness centrality \cite{Freeman1979}, and eigenvector centrality \cite{Bonacich2007}. Degree centrality is a straightforward and efficient metric but less relevant, because a node having a few highly influential neighbors may be more influential than a node having a larger number of less influential neighbors, while the computation of closeness, betweenness and eigenvector centrality is highly time-consuming and thus usually not feasible for large-scale networks. Some methods have already been proposed in the literatures to identify influential spreaders in an effective and efficient way \cite{Valery2008,Maksim2010,Cesar2011,Bonan2012,Frank2012,Chen2012,Chen2013,An2013,Liu2013}. Kitsak \emph{et al.} \cite{Maksim2010} proposed a coarse-grained method by using $k$-core decomposition to quantify a node's influence based on the assumption that nodes in the same shell have similar influence and nodes in higher-level shells are likely to infect more nodes. Zeng \emph{et al.} \cite{An2013} found several limitations of the above method and proposed a mixed degree decomposition method, as an improved version of the traditional $k$-core decomposition. Liu \emph{et al.} \cite{Liu2013} argued that nodes in the same shell may indeed have considerably different spreadabilities, and proposed an improved method that can properly rank nodes in the same shell. Chen \emph{et al.} \cite{Chen2012} devised a semi-local index by considering the next nearest neighborhood, which performs as good as global indices while has much lower computational complexity, and thus obtains a good trade-off on effectiveness and efficiency. In addition, Chen algorithm \cite{Chen2012} can well identify influential nodes in a hierarchical tree that cannot be managed by the $k$-core decomposition. Furthermore, Chen \emph{et al.} \cite{Chen2013} proposed an improved index for directed networks, which accounts for the effects of local clustering and shows better performance. 

Recently, L\"u \emph{et al.} \cite{L2011} proposed the LeaderRank algorithm to identify influential spreaders in directed networks, which is a simple variant of PageRank \cite{Brin1998}, namely a so-called ground node connected with every other node by a bidirectional link is introduced into the original network, and then the standard random walk process is applied to dig out influential spreaders. Thanks to this simple change, LeaderRank outperforms PageRank in several aspects: (i) LeaderRank converges faster since the network is strongly connected with diameter being only 2; (ii) the influential nodes identified by LeaderRank can spread information faster and wider than those by PageRank; (iii) LeaderRank has higher tolerance to noisy data, and (iv) LeaderRank has higher robustness to intentional attacks. In this paper, we further improve the LeaderRank algorithm by allowing nodes with more fans get more scores from the ground node, that is, replacing the standard random walk by a biased random walk. Experiments on three real social networks (i.e., Delicious, Epinions and Slashdot) show that the weighted LeaderRank can considerably improve the performance of original LeaderRank.

\section{Algorithm}

Given a network consisting of $N$ nodes and $M$ directed links, a ground node connected with every node by a bidirectional link is added. Then, the network becomes strongly connected and consists of $N+1$ nodes and $M+2N$ links (a bidirectional link is counted as two links with inverse directions). LeaderRank directly applies the standard random walk process to determine the score of every node. Accordingly, if the score of node $i$ at time step $t$ is $s_i(t)$, the dynamics can be described by an iterative process as
\begin{equation}
s_i(t+1)=\sum_{j=1}^{N+1}\frac{a_{ji}}{k_j^{out}}s_j(t),
\end{equation}
where $a_{ji}$ is the element of the corresponding $(N+1)$-dimensional adjacency matrix, which equals 1 if there is a directed link from $j$ to $i$ and 0 otherwise, and $k_j^{out}$ is the out-degree of node $j$. The process starts with the initialization where all node scores are 1 and will soon converge to a unique steady state denoted as $s_i^\infty$ ($i=1,2,\cdots,N,N+1$). LeaderRank ranks all nodes according to $s_i^\infty$, and the nodes with larger final scores are considered to be more influential in spreading.

\begin{figure}
\begin{center}
\scalebox{0.45}[0.45]{\includegraphics{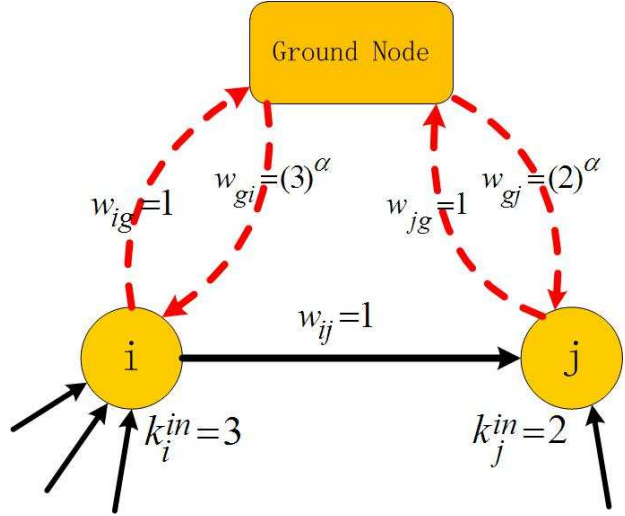}}
\caption{Illustration of the weighted LeaderRank algorithm, where a user having more fans would get more score from the ground node $g$.
The weight of the directed link from the ground node to node $i$ is $w_{gi}=(k_i^{in})^\alpha$, and other weights are all one. }\label{Fig1}
\end{center}
\end{figure}

As we have mentioned in the Introduction, LeaderRank outperforms PageRank in several aspects. In this paper, we intend to further improve LeaderRank by introducing a weighted mechanism. Considering a social network where a user $i$ is called a fan of user $j$ if there is a directed link from $i$ to $j$, namely $i$ could receive information from $j$ and thus $j$ will receive scores from $i$ (if a node's fans are of high influence, this node will be highly influential as well). Obviously, the number of fans (i.e., in-degree) is an important local indicator for a user's influence in spreading. Therefore, based on LeaderRank, we allow nodes with different in-degrees get different scores from the ground node (other possible weighting schemes will be discussed in the last section). Accordingly, the network is described by an $(N+1)$-dimensional weighted adjacency matrix $W$. As illustrated in Fig. 1: (i) if $a_{ij}=0$, then $w_{ij}=0$; (ii) for any normal node $i$ and the ground node $g$, $w_{gi}=(k_i^{in})^\alpha$ and $w_{ig}=1$, where $\alpha$ is a free parameter; (iii) for all other cases, $w_{ij}=1$. After determining the weight of every link, the dynamics follows a biased random walk \cite{Ou2007}, namely the score from node $j$ to node $i$ is proportional to the weight $w_{ji}$:
\begin{equation}
s_i(t+1)=\sum^{N+1}_{j=1}\frac{w_{ji}}{\sum^{N+1}_{l=1}{w_{jl}}}s_j(t).
\end{equation}
Same to LeaderRank, we use final scores in the steady state to quantify nodes' influences.

\begin{figure}
\begin{center}
\scalebox{0.7}[0.7]{\includegraphics{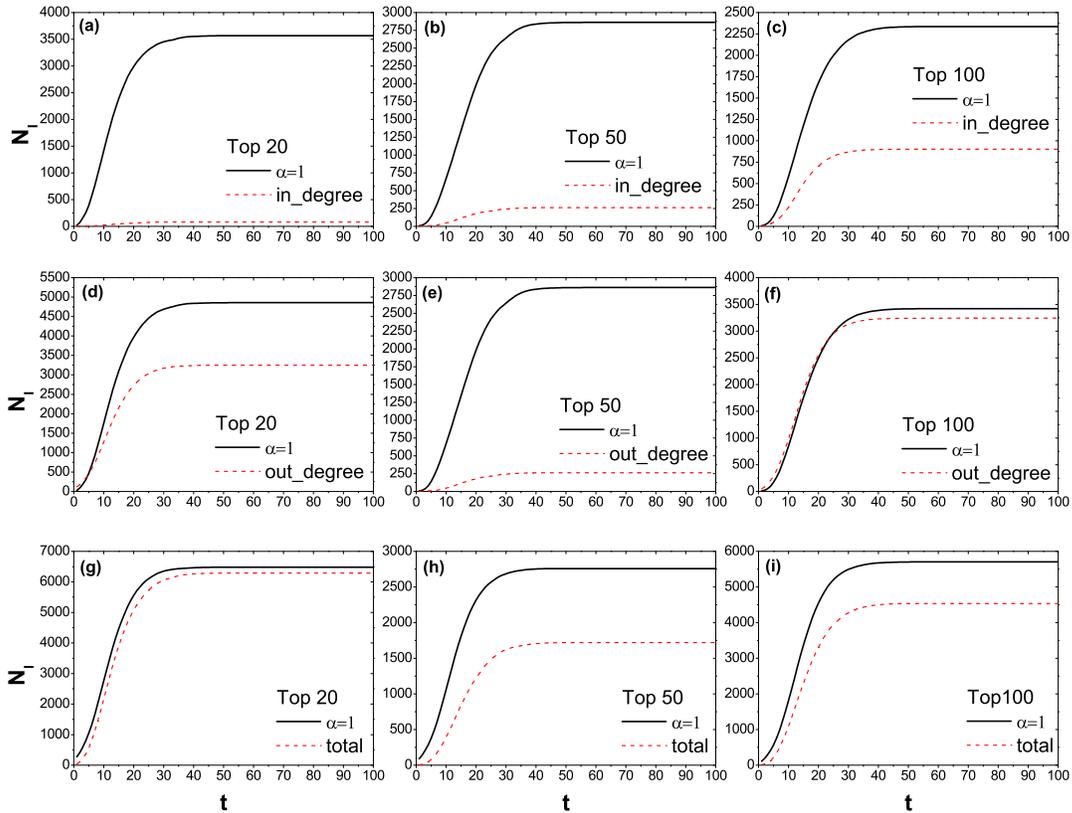}}
\caption{The number of nodes having been infected as a function of time for Delicious. The three columns, from left to right, are respectively for $L=20$, $L=50$ and $L=100$, while the three cows, from up to down, are respectively for in-degree, out-degree and total degree. The black and red curves denote the results from weighted LeaderRank and degree centralities, respectively. The parameters are set as $\alpha=1$ and $\beta=0.06$, and the results are obtained by averaging over 100 independent runs. }
\end{center}
\end{figure}

\section{Results}

\subsection{Data}

To validate the effectiveness of the weighted LeaderRank algorithm, we test it on three real social networks: (i) Delicious \cite{data1}---A social bookmarking
web site, where users could subscribe to leaders' collections as
their sources of information; (ii) Epinions \cite{data2}---A product
review web site, where users could select to trust or distrust the
reviews written by others; (iii) Slashdot \cite{data3}---A technological
news web site, which allows users tag others as friends or
foes. Basic statistics of these three real networks are presented in Table 1.

\begin{table}[htbp]
\begin{center}
\caption{\label{comparison} Basic statistics of three real networks, including the number of nodes $N$, the number of links $M$ and the average degree $\langle k\rangle = \langle k^{in}\rangle =\langle k^{out}\rangle$. }
\begin{tabular}{llll}
\hline
\hline
Networks     & $N$   & $M$    & $\langle k\rangle$   \\
\hline
delicious &582377 &1686131  &2.90  \\
\hline
Epinions &75879 &508837 &6.71 \\
\hline
Slashdot     &77360 &828161 &10.71 \\
\hline
\hline
\end{tabular}
\end{center}
\end{table}

\subsection{Comparison with Degree Centralities} 

We employ the standard susceptible-infected-removed (SIR) model \cite{Anderson1992} to estimate the spreading influence of
the top-ranked users. In the SIR model, each node can be in one of the three possible states: susceptible, infected and removed. Initially, a single node is set to be infected, and at each time step, each infected node will infect all its susceptible neighbors with
probability $\beta$ and then be removed (dead or recovered with immunity) with probability $\gamma$. Without the loss of generality, we set $\gamma=1$. The dynamical process stops when no infected node is present.

We first compare the spreading processes activated by top-ranked nodes from weighted LeaderRank and the traditional degree centralities. There are three degree centralities for directed networks: in-degree $k^{in}$, out-degree $k^{out}$ and total degree $k^{tot}=k^{in}+k^{out}$. Taking in-degree as an example. If among the two top-$L$ lists by weighted LeaderRank and in-degree, there are $n$ different nodes, we compare the spreading processed activated by these $n$ different nodes, respectively. For example, if $L=4$ and for weighted LeaderRank and in-degree, the top-ranked lists are $\{7,28,392,1146\}$ and $\{28,168,433,1146\}$, then $n=2$ and we will compare the spreading processes with initially infected nodes being $\{7,392\}$ for weighted LeaderRank and $\{168,433\}$ for in-degree, respectively. In the comparison, we fix $\alpha=1$ since this value is close to the optimal value for all three real networks (see later). Notice that, when $\beta$ is very small, the disease cannot spread out and the comparison is less meaningful, while when $\beta$ is very large, almost every individual will get infected, and thus the advantage of our algorithm is not obvious. Only when $\beta$ lies in the middle, the comparison is meaningful. In the following simulations, we always set the values of $\beta$ as 0.06, 0.015 and 0.015 for Delicious, Epinions and Slashdot, respectively. We have checked that the qualitative results are not sensitive to $\beta$ unless $\beta$ is very large or very small. 
 
Figures 2, 3 and 4 respectively report the number of nodes, $N_I$, having been infected (i.e., infected and removed nodes) as a function of time for Delicious, Epinions and Slashdot. In each figure, we compare the weighted LeaderRank with three degree centralities for three typical lengths of top-ranked lists: $L=20$, $L=50$ and $L=100$. Obviously, in each of the 27 reported cases, the weighted LeaderRank at $\alpha=1$ outperforms degree centralities.

\begin{figure}
\begin{center}
\scalebox{0.7}[0.7]{\includegraphics{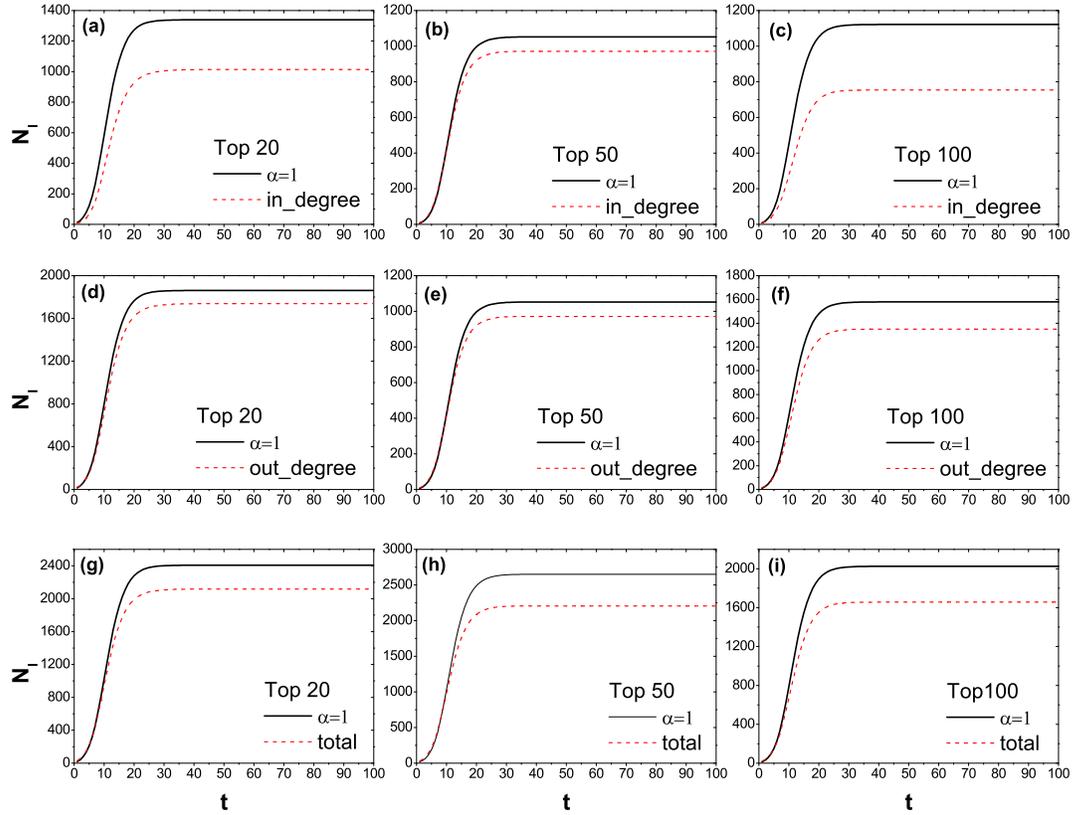}}
\caption{The number of nodes having been infected as a function of time for Epinions. The three columns, from left to right, are respectively for $L=20$, $L=50$ and $L=100$, while the three cows, from up to down, are respectively for in-degree, out-degree and total degree. The black and red curves denote the results from weighted LeaderRank and degree centralities, respectively. The parameters are set as $\alpha=1$ and $\beta=0.015$, and the results are obtained by averaging over 100 independent runs. }
\end{center}
\end{figure}

\begin{figure}
\begin{center}
\scalebox{0.7}[0.7]{\includegraphics{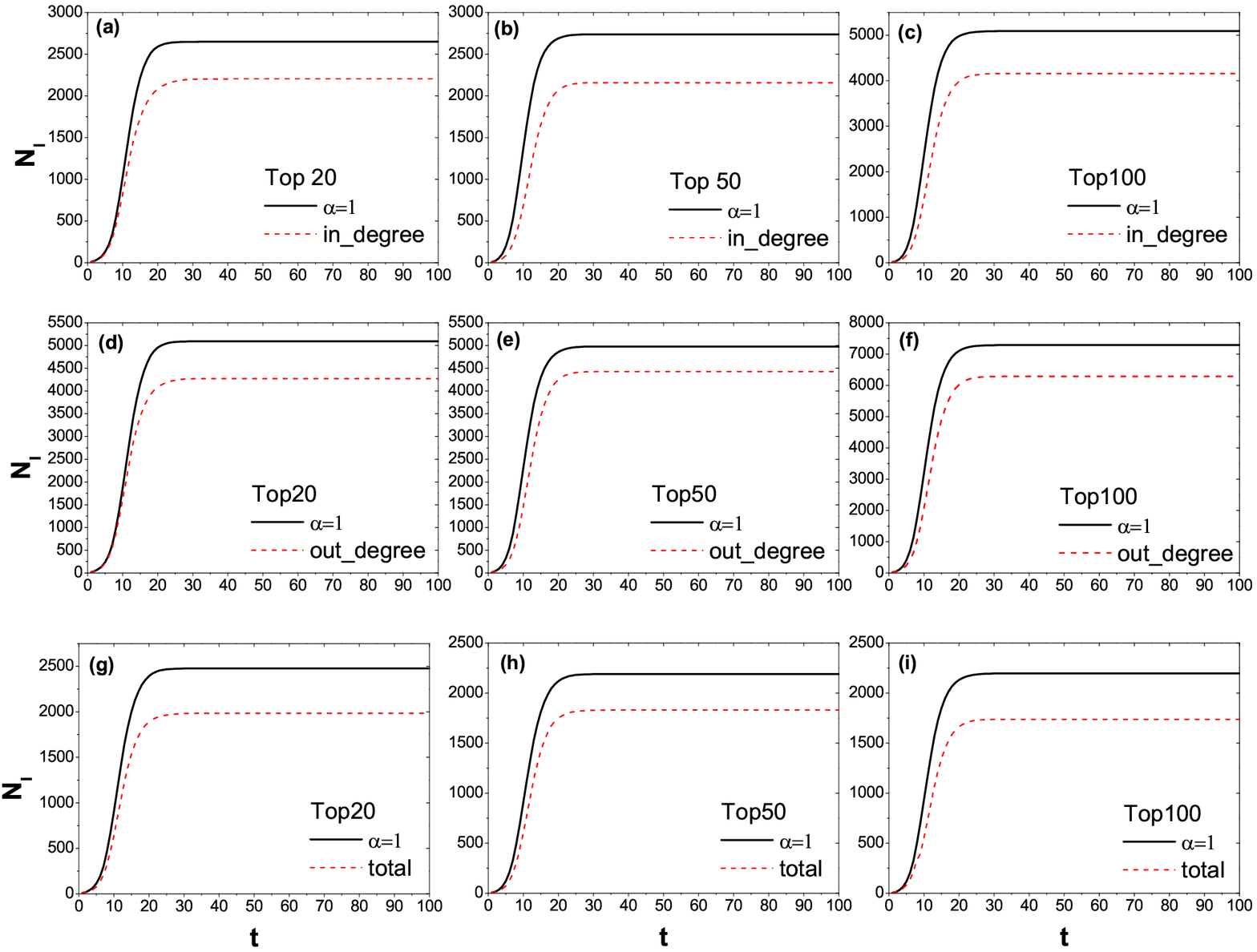}}
\caption{The number of nodes having been infected as a function of time for Slashdot. The three columns, from left to right, are respectively for $L=20$, $L=50$ and $L=100$, while the three cows, from up to down, are respectively for in-degree, out-degree and total degree. The black and red curves denote the results from weighted LeaderRank and degree centralities, respectively. The parameters are set as $\alpha=1$ and $\beta=0.015$, and the results are obtained by averaging over 100 independent runs. }
\end{center}
\end{figure}

\subsection{Difference between LeaderRank and Weighted LeaderRank}

Note that, when $\alpha=0$, the weighted LeaderRank degenerates to the original LeaderRank. Considering the top-$L$ influential nodes respectively produced by LeaderRank and weighted LeaderRank given a specific $\alpha$, if there are $n$ different nodes in the two lists, we define the \emph{difference} of such two algorithms as $n/L$. Figure 5 reports the difference between LeaderRank and Weighted LeaderRank for three real networks, indicating that the top-$L$ list is sensitive to the parameter $\alpha$. Generally speaking, the larger the absolute value of $\alpha$ is, the bigger the difference is.

\begin{figure}
\begin{center}
\scalebox{0.7}[0.7]{\includegraphics{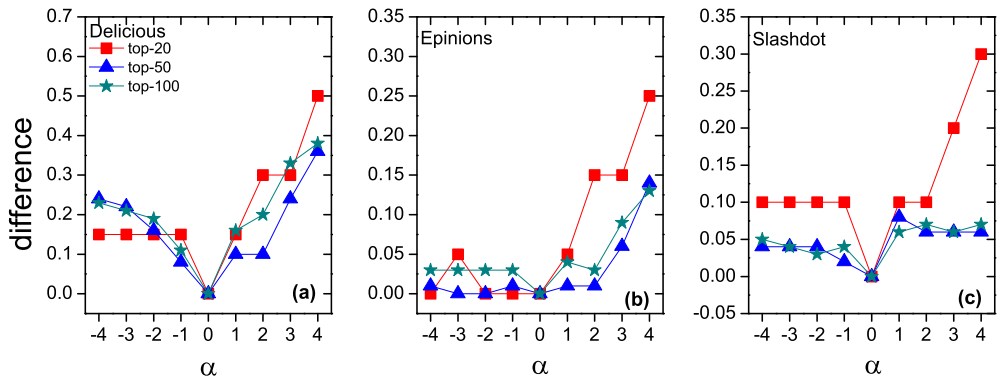}}
\caption{The fraction of different nodes in top-$L$ ($L=20,50,100$) list of influential nodes between LeaderRank ($\alpha$=0) and weighted LeaderRank. (a), (b) and (c) display the results for Delicious, Epinions and Slashdot, respectively. }\label{Fig2}
 \end{center}
\end{figure}

\subsection{Spreadability}

Given the network, the spreadability of a node $i$, denoted by $SP_i$, is defined as the average number of removed nodes at the steady state if node $i$ is set to be the only initially infected node. In this paper, the spreadability of a given node is obtained by averaging over 100 independent runs.

Given the top-ranked lists by the LeaderRank and the weighted LeaderRank with parameter $\alpha$, if there are $n$ different nodes, then the average values of spreadability of the $n$ nodes for the weighted LeaderRank and LeaderRank are respectively denoted by $\overline{SP}(\alpha)$ and $\overline{SP}(0)$. Accordingly, we define the relative spreadability as
\begin{equation}
\eta=\frac{\overline{SP}(\alpha)-\overline{SP}(0)}{\overline{SP}(0)}.
\end{equation}
Clearly, $\eta>0$ indicates that the weighted LeaderRank performs better than the LeaderRank.

\begin{figure}
\begin{center} \scalebox{0.7}[0.7]{\includegraphics{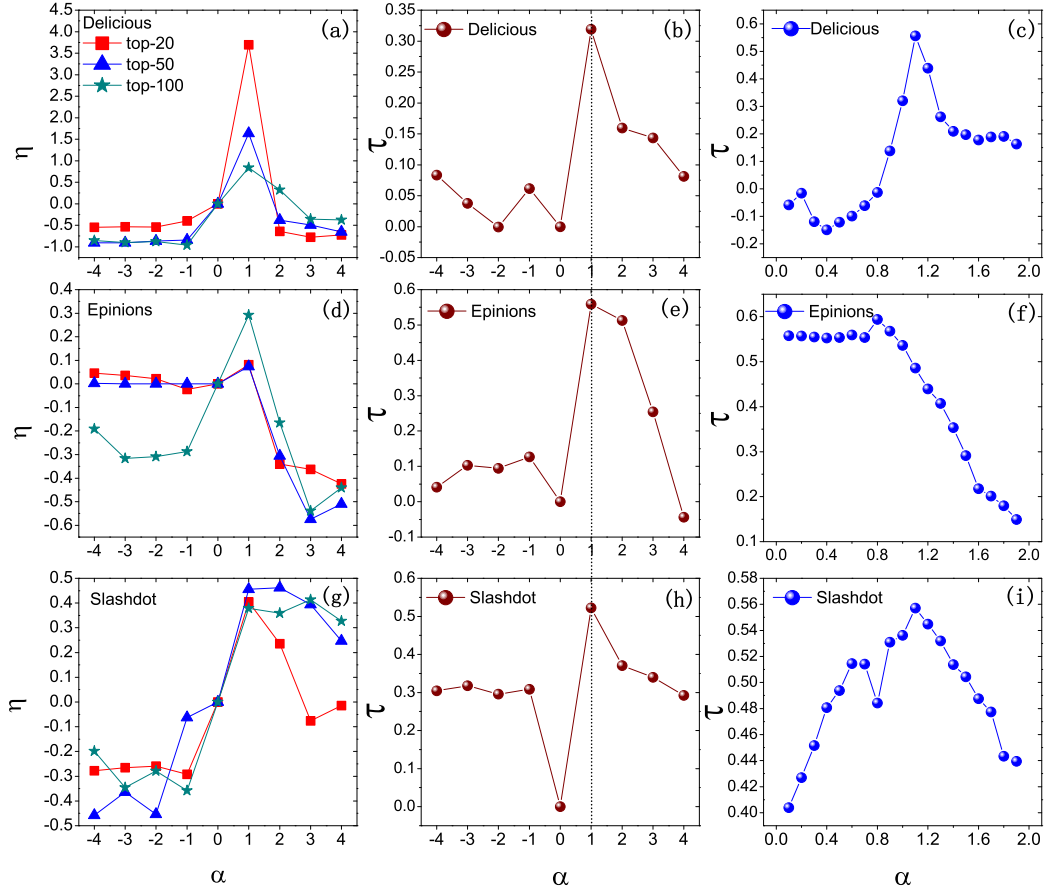}}
\caption{Relative spreadability at $L=20$, 50 and 100 for (a) Delicious, (d) Epinions, and (g) Slashdot, with parameters being $\beta=0.06$, 0.015 and 0.015, respectively. The correlation between the change of nodes' scores and their spreadabilities, quantified by the Kendall's Tau, for (b)(c) Delicious, (e)(f) Epinions, and (h)(i) Slashdot.}\label{Fig3}
\end{center}
\end{figure}

Figure 6 shows the relative spreadability, which demonstrates that the weighted LeaderRank can considerably improve the spreadability of top-ranked nodes compared with the original LeaderRank, and the optimal performance is around $\alpha=1$ for the three present cases. Denote $s_i^\infty(\alpha)$ the final score of node $i$ by the weighted LeaderRank with parameter $\alpha$, then the change of $i$'s score from the original LeaderRank reads
\begin{equation}
\Delta_{i}=s_{i}^\infty(\alpha)-s_{i}^\infty(0).
\end{equation}
We quantify the correlation between score changes and spreadabilities by the well-known Kendall's Tau \cite{Kendall1938}. For a pair of nodes $i$ and $j$, if $SP_i>SP_j$ and $\Delta_i>\Delta_j$ or $SP_i<SP_j$ and $\Delta_i<\Delta_j$, we call this pair is a concordant pair, otherwise it is called a discordant pair. If the numbers of concordant pair and discordant pair are $M_{con}$ and $M_{dis}$, then the Kendall's Tau is
\begin{equation}
\tau=\frac{M_{con}-M_{dis}}{\frac{1}{2}N(N-1)}.
\end{equation}
$\tau>0$ indicates a positive correlation and $\tau=1$ is the ideal case where $SP_i$ and $\Delta_i$ are completely in the same order. As shown in figure 6(b), 6(e) and 6(h), the peaks of Kendall's Tau are also around $\alpha=1$, in accordance with the optimal $\alpha$ for $\eta$. Results with higher resolution, as shown in figure 6(c), 6(f) and 6(i) show the optimal values are $\alpha=1.1$, $\alpha=0.8$ and $\alpha=1.1$, respectively.

\subsection{Robustness}
\begin{figure}
\begin{center} \scalebox{0.7}[0.7]{\includegraphics{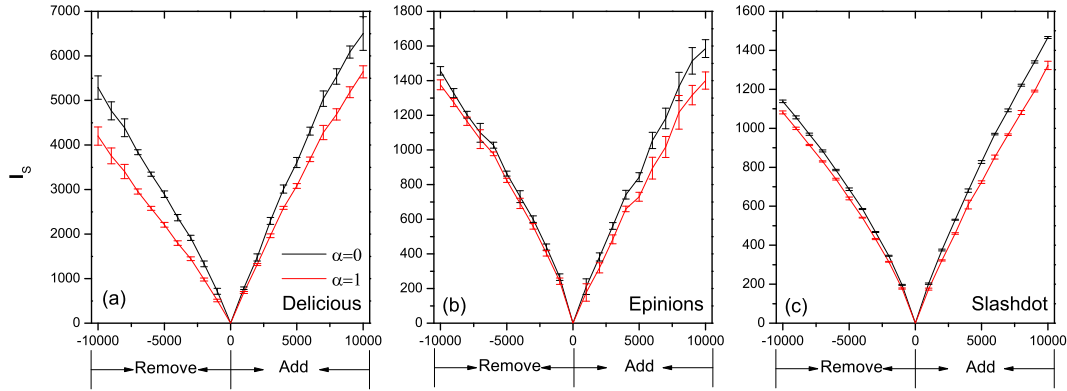}}
\caption{The change of scores, $I_s$, as a function of the number of links
being randomly added or removed for (a) Delicious, (b) Epinions, and (c) Slashdot.}\label{Fig5}
\end{center}
\end{figure}

Real networks are usually noisy and thus many efforts have been made to detect missing links and remove spurious links \cite{Lu2011b}. Therefore, a good ranking algorithm should be tolerant to noisy data. Here we randomly add or remove some links from the original networks to see the change of scores, as
\begin{equation}
I_s=\sum_{i=1}^{N}|\tilde{s}_{i}^\infty-s_{i}^\infty|,
\end{equation}
where $\tilde{s}_{i}^\infty$ is the final score of node $i$ after adding or removing a certain number of links. Of course, $I_s$ increases as the increasing of randomly adding or removing links, as shown in figure 7. While comparing with the original LeaderRank, the weighted LeaderRank with $\alpha=1$ is more tolerant, as indicated by its smaller $I_s$.

Malicious activities are common in the Internet, in particular in social networks and e-commerce user-product bipartite networks when users manipulate to gain skewed reputation \cite{Masum2004}, and thus to design robust ranking algorithms against spammers attracts increasing attention recently \cite{Zhou2011}. Here we consider a representative example called \emph{sybil attack} \cite{Levine2006}, in which spammers deliberately create fake entities to obtain disproportionately high rank. Under the Sybil attack, a spam user $i$, with original rank being denoted by $r_i$, creates $v$ fake fans all pointing to $i$ itself. Then $i$¡¯s new rank is $r¡¯_i$. Clearly, $r¡¯_i$ should be smaller than $r_i$. The smaller difference $|r'_i - r_i|$ corresponds to higher robustness.

Specifically, each time we consider a spam user who creates $v=20$, $50$ or $100$ fake fans and compare the manipulated ranks of this user by LeaderRank and weighted LeaderRank with $\alpha=1$. Only the top-100 users (i.e., $r_i\in \{1,2,\cdots,100\}$ are under investigation. Figure 8 reports the simulation results, where the horizontal axis shows the original rank of a user, and the vertical axis shows her manipulated rank after the addition of $v$ fake fans. As shown in figure 8, the weighted LeaderRank with $\alpha=1$ is more robust against sybil attack as the change of ranks is much smaller than that by LeaderRank.

\begin{figure}
\begin{center} \scalebox{0.7}[0.7]{\includegraphics{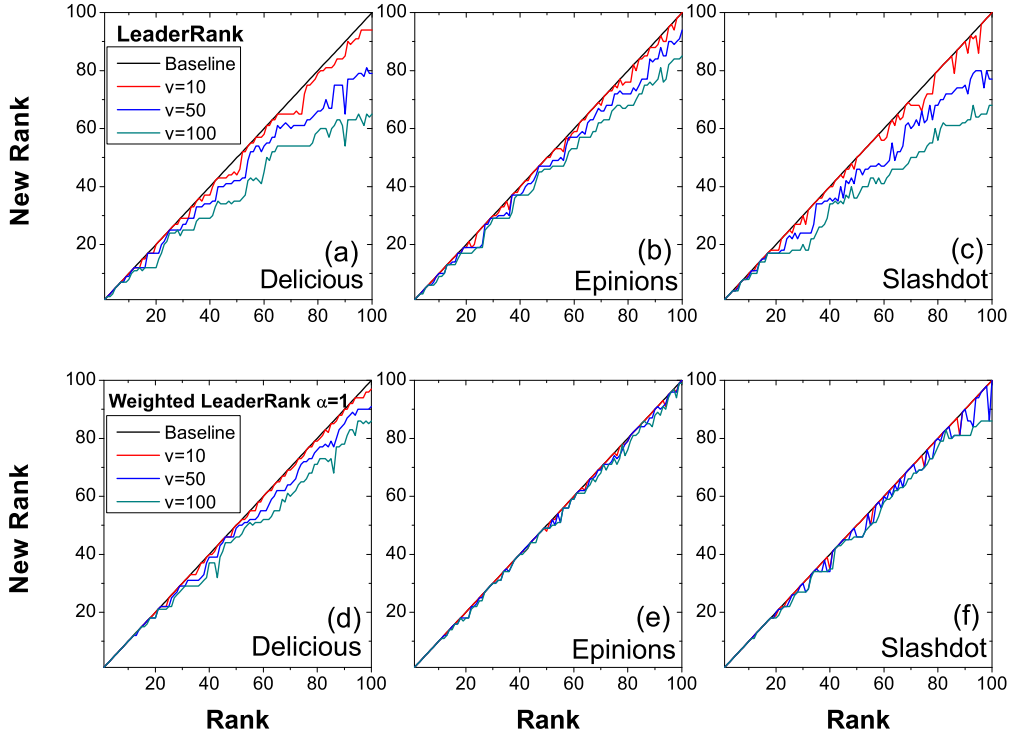}}
\caption{The manipulated rank as obtained by LeaderRank and weighted LeaderRank with
$\alpha=1$, after the addition of $v$ fake fans ($v=10,50,100$) for (a)(d) Delicious, (b)(e) Epinions, and (c)(f) Slashdot.}\label{Fig6}
\end{center}
\end{figure}

\section{Conclusion and Discussion}
LeaderRank may serve as a prototype of ranking algorithms applicable to rank users in social networks. Compared with PageRank, it can find higher influential spreaders with faster convergence, and is more robust to noise and spammers. In this paper, we further improve the LeaderRank by allowing nodes with more fans get more scores from the ground node. With almost the same converging speed (we have checked by simulations), this so-called weighted LeaderRank performs better than LeaderRank in three aspects: (i) the ability to find out more influential spreaders, (ii) the higher tolerance to noisy data, and (iii) the higher robustness to intentional attacks.


Since the in-degree of a node (i.e., the number of fans) directly indicates its influence, it is natural to weight nodes according to their influences as applied in this paper. In contrast, it is strange to weight a node according to how many nodes it follows (i.e., the number of its leaders, namely its out-degree). In addition, to weight a node by the number of its leaders is not robust since a node is easy to add as many as leaders. We can also set $w_{gi}=1$ and $w_{ig}=(k_i^{in})^\alpha$. However, it is also not easy to explain: why a node with more fans need to give more to the ground node? We have checked that all the three above-mentioned weighting schemes give worse performance than the present one. Using two or more parameters in a more complicated form can give a slightly better performance, but it is very ugly and not easy to be understood.

\section*{Acknowledgement}
This work is supported by the National Natural Science Foundation
of China (Grant Nos. 11165003, 11205042 and 11222543), the Program for Excellent
Talents in Guangxi Higher Education Institutions, the Huawei university-enterprise
cooperation project YBCB2011057. and the Program for New Century Excellent Talents in
University under Grant No. NCET-11-0070.

\end{document}